\documentclass[preprint]{aastex}
\usepackage{amsmath}

\newcommand*{\rom}[1]{\expandafter\@slowromancap\romannumeral #1@}

\slugcomment{}
\title{Microlensing Constraints on Broad Absorption and Emission Line Flows in the Quasar H1413+117}

\shorttitle{Microlensing Constraints on H1413+117}
\shortauthors{O'Dowd et al.}

\begin{document}


%\title{}

\author{Matthew J. O'Dowd\altaffilmark{1}\altaffilmark{2}\altaffilmark{3}, Nicholas F. Bate\altaffilmark{4}\altaffilmark{5}, Rachel L. Webster\altaffilmark{4}, Kathleen Labrie\altaffilmark{6}, and Joshua Rogers\altaffilmark{1}}

\altaffiltext{1}{Department of Physics and Astronomy, Lehman College of the CUNY, Bronx, NY 10468, USA}
\altaffiltext{2}{Department of Astrophysics, American Museum of Natural History, Central Park West and 79th Street, NY 10024-5192, USA}
\altaffiltext{3}{The Graduate Center of the City University of New York, 365 Fifth Avenue, New York, NY 10016, USA}
\altaffiltext{4}{School of Physics, University of Melbourne, Parkville, Victoria 3010, Australia}
\altaffiltext{5}{Sydney Institute for Astronomy, School of Physics, A28, University of Sydney, NSW, 2006, Australia}
\altaffiltext{6}{Gemini Observatory, Hilo, HI 96720, USA}

\begin{abstract}
We present new integral field spectroscopy of the gravitationally lensed broad absorption line (BAL) quasar H1413+117, covering the ultraviolet to visible rest-frame spectral range. We observe strong microlensing signatures in lensed image D, and we use this microlensing to simultaneously constrain both the broad emission and broad absorption line gas. By modeling the lens system over the range of probable lensing galaxy redshifts and using on a new argument based on the wavelength-independence of the broad line lensing magnifications, we determine that there is no significant broad line emission from smaller than $\sim$20 light days. We also perform spectral decomposition to derive the intrinsic broad emission line (BEL) and continuum spectrum, subject to BAL absorption. We also reconstruct the intrinsic BAL absorption profile, whose features allow us to constrain outflow kinematics in the context of a disk-wind model. We find a very sharp, blueshifted onset of absorption of 1,500~km/s in both C{\sc iv} and N{\sc v} that may correspond to an inner edge of a disk-wind's radial outflow. The lower ionization Si{\sc iv} and Al{\sc iii} have higher-velocity absorption onsets, consistent with a decreasing ionization parameter with radius in an accelerating outflow. There is evidence of strong absorption in the BEL component which indicates a high covering factor for absorption over two orders of magnitude in outflow radius.
\end{abstract}

\bigskip
\bigskip

\section{Introduction}

The central engines of powerful Type I active galactic nuclei (AGNs) contain gas that is moving at extreme velocities, apparent from their broad emission lines (BELs) and, in the case of broad absorption line (BAL) quasars, in blueshifted absorption troughs. These spectral features arise from compact gas flows close to the accretion disk, apparent from the presence of high ionization lines and size scales from reverberation mapping  \citep{Blandford82, Peterson93, Kaspi07}.
These compact flows are probably the origin of larger-scale outflows (e.g. Chelouche \& Netzer 2005; Crenshaw \& Kraemer 2007; Storchi-Bergmann 2010) and result in feedback that is believed to be important on many scales, 
regulating supermassive black hole (SMBH) fueling, quenching star formation, and so driving galaxy-SMBH coevolution (e.g. Silk and Rees 1998; Fabian 1999; Sommerville et al. 2008; Zubovas \& King 2012)
as well as influencing the intergalactic and intracluster medium and affecting galaxy and large-scale structure formation (e.g. Scannapieco \& Oh 2004; Rafferty, McNamara \& Nulsen 2006; Sommerville  et al. 2008; Fabjan  et al. 2010; Wang 2015).
Yet despite the importance of these processes, the connection between compact and large-scale outflows is not understood. In the case of the broad emission line region (BELR), even the basic kinematics are unclear; independent reverberation mapping results support outflow, inflow, and orbital motion \citep{Kaspi00, Kaspi07, Denney09, Bentz09, Doroshenko12}, and microlensing results have been consistent with both virial BELRs \citep{ODowd11, Sluse11} and with more complex, multi-component flows \citep{Richards04, Guerras13}.

BAL quasars offer perhaps the clearest evidence of nuclear outflows in AGNs. 
Approximately 15\% of optically selected quasars exhibit high velocity, blueshifted absorption features \citep{Reichard03}. The outflows responsible for these features may be common to most quasars if this 15\% represents a special line of sight to outflows with a similar covering factor \citep{Weymann91, Gallagher07}, or a particular phase in the lifecycle of quasars.
An accurate measurement of the absorption profile of a BAL quasar would go a long way to revealing the kinematics of these outflows, however this measurement is complicated by the fact that their intrinsic absorption profile cannot be recovered without knowing the shape of the underlying BEL that it absorbs. These BEL profiles can vary dramatically between quasars, and in the case of BAL quasars may be strongly affected by a biased viewing angle. 

The gravitationally lensed BAL quasar H1413+117 (the Cloverleaf; Magain et al. 1988) offers a unique opportunity to simultaneously study both BEL and BAL flows in the same source. The four lensed images of this $z=2.55$ quasar exhibit deep BAL features in high-ionization lines, as well as low-ionization absorption in Al{\sc iii}, making it a LoBAL quasar. Significant microlensing has been observed in H1413+117 over the past 15 years \citep{Angonin90, Ostensen97, Monier98, Anguita08, Huts}. In microlensing, substructure within the foreground lensing galaxy causes changes in the magnification of individual lensed images. The degree of microlensing depends on size of lensed region, and in the case of H1413+117 this has resulted enhanced magnification in the continuum in lensed image D compared to the broad emission lines in this component. 

Given a model of the microlensing magnification of the lens system, these differential magnifications can be used to derive size constraints of emission regions. This approach has been used to contrain accretion disk size and temperature gradient in several lensed quasars (e.g. Kochanek et al. 2004; Bate et al. 2008; Floyd, Bate \& Webster 2009; Blackburne et al. 2011; Jim\'enez-Vicente et al. 2014; Blackburne et al. 2015), including in H1413+117 by \citet{Huts} (hereafter H10). It has also provided constraints on BELR size \citep{Kochanek04, Richards04, Wayth05, Sluse07, Guerras13} and kinematics \citep{Richards04, ODowd11, Sluse11, Guerras13} in a number of quasars, however no such constraints have been derived for the BELR of H1413+117.

Another novel use of microlensing takes advantage of differential magnifications between lensed images to algebraically decompose the spectral contributions of different physical regions within the quasar \citep{Sluse07}. \citet{Angonin90} (hereafter A90) and H10 apply this technique to H1413+117, shedding light on the true shape of the broad absorption profile, and in the case of H10 decomposing into microlensed and non-microlensed components which reveal substructure in each component and may point to a two-component outflow comprising a polar outflow and equatorial disk-wind.

In this paper we present new Gemini North GMOS integral field spectroscopy of H1413+117, and based on the strong microlensing observed in lensed image D we derive new statistical size constraints on the BELR. We also present a new spectral decomposition, deriving BELR and continuum components, as well as the intrinsic BAL profiles for N{\sc v}, Si{\sc iv}, C{\sc iv}, and Al{\sc iii}. 

In Section \ref{obs} we describe the observations and extraction of lensed image spectra. In Section \ref{sigs} we analyse the microlensing signatures in this new epoch of data. In Section \ref{microlensingbehavior} we use lensing simulations to explore the size dependence of the observed microlensing signatures and deriving size constraints for the BELR. 
In Section \ref{algebra} we decompose the spectra in terms of the broad emission line region and continuum, emphasizing the physical interpretation of the decomposed regions, and we derive the intrinsic BAL absorption profile. We discuss the interpretation of these decomposed components in the context of disk-wind outflow models in Section \ref{disc}. In Section \ref{conc} we summarize our conclusions.

\section{Observations and Spectral Extraction}
\label{obs}

\begin{figure}[ht]
\includegraphics*[width=80mm]{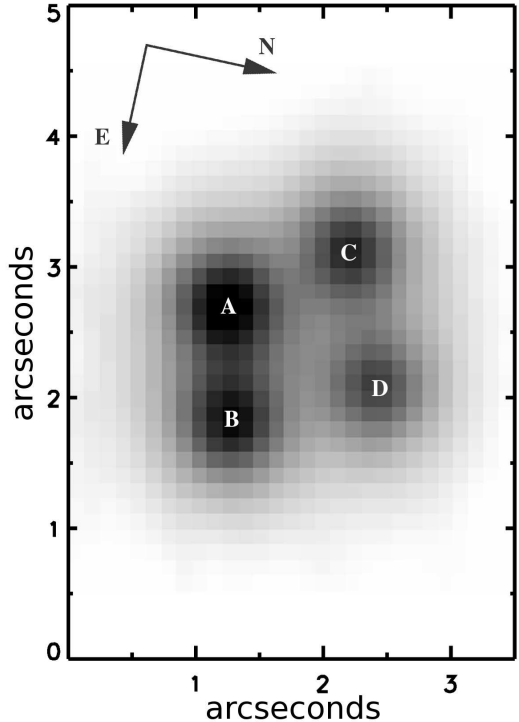}
  \caption{2-D image created from collapsed IFU data
    cube. Letters indicate the standard designation of quasar lensed images in H1413$+$117.}
  \label{2dimage}
\end{figure}

H1413$+$117 was observed on the night of April 5th, 2006 
with the GMOS Integral Field Unit (IFU)
\citep{AllingtonSmith, Hook} on the Gemini North telescope (program ID:
GN-2006A-Q-15). 
3$\times$20 minute exposures were taken in one-slit
mode using the B600 grating with 
a spectral range of 3961\AA\ to 6845\AA\ and a resolution of $R=1688$.
At the redshift of H1413$+$117, $z=2.55$, this covers the Ly$\alpha$ through C{\sc iii]}
broad emission lines.
Seeing conditions were good, with a PSF FWHM of 0\farcs8.

\subsection{Data Reduction}

The data was reduced using the Gemini IRAF package v1.10 under IRAF 2.14.1.  We used calibration files taken as part of the standard Gemini Facility Calibration Unit (GCAL) suite.  Mostly, the standard GMOS IFU reduction procedure was followed for overscan subtraction and trimming, bias subtraction, flat fielding and response correction, arc calibration, fiber extraction, sky subtraction, flux calibration, cube reconstruction with atmospheric dispersion correction.

Additionally, we corrected the GCAL flat and the twilight flat for scattered light by fitting a surface to residual light detected between the fiber bundles.  That correction was not necessary for the science observations, the signal being too weak to generate any significant scattered light.  The cosmic rays were corrected before extraction with LA-Cosmic \citep{vanDokkum01}, followed, after extraction, by a floor and ceiling masking of remaining extreme hits.  The calibrated cubes for each science observation were aligned and combined with a customized version of the pyfu software package\footnotemark[1].

\footnotetext[1]{Scattered light correction tasks by Bryan Miller, Gemini Observatory (version 2010).  LA-Cosmic IRAF wrapper and pyfu suite by James Turner, Gemini Observatory (version 2010).  All, except the pyfu software, are now part of the public Gemini IRAF package (v1.13).  The pyfu software is available on the Gemini Data Reduction Forum at http://drforum.gemini.edu/topic/pyfu-datacube-mosaicking-package/}

Figure \ref{2dimage} shows the reduced data cube collapsed into
a 2-D image. All lensed components are distinct and well-resolved.

\subsection{Extraction of Image Spectra}
\label{psffits}

\begin{figure*}[ht]
\includegraphics[width=160mm]{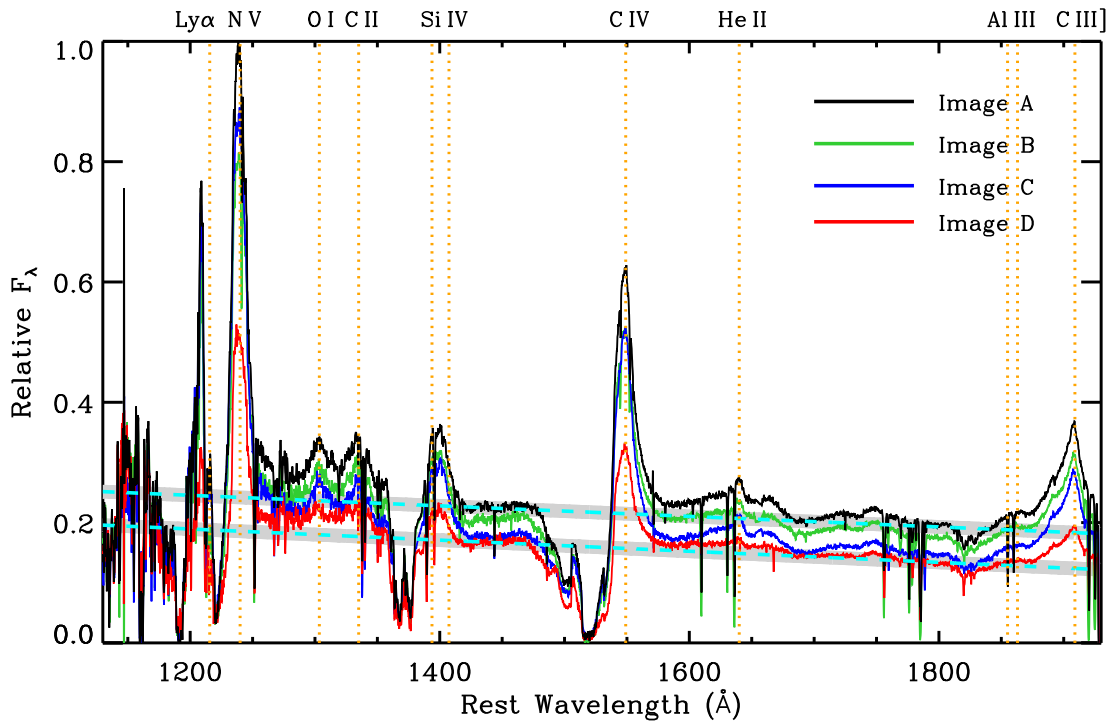}
  \caption{Extracted spectra for the four quasar lensed components:
  black: image A, green: image B, blue: image C, red: image D. Vertical dotted
  lines mark the positions of emission lines assuming $z=2.55$. Dashed lines indicate the best linear fits to the continuum spectra of lensed image A \& D.}
 \label{spec}
\end{figure*}

The lensed images are well resolved in these observations, allowing for a good extraction of their spectra. 
We perform this extraction using the method described in \citet{ODowd11}. We
refer the reader to that paper for a detailed description, however in
short: relative lensed image fluxes at each wavelength bin are
determined by least-squares fitting of reconstructed Point Spread
Functions (PSFs). A separate PSF is determined for each 100\AA-wide spectral bin.
Each given PSF is initially estimated using carefully-masked fragments of
the IFU frame (stacked over 100\AA), and
are iteratively improved  with a series of least-squares minimizations
of each frame.

Figure~\ref{spec} shows the extracted spectra of the four quasar images. 
We see distinct broad emission lines for Ly$\alpha$, N{\sc v}, O{\sc i}, C{\sc ii}, Si{\sc iv}, C{\sc iv}, and C{\sc iii]} as well as prominent iron emisison features at $\sim$1600-1800\AA. Emission from He{\sc II} and Al{\sc iii} is also apparent, blended with iron and C{\sc iii]} respectively. Clear redshifted broad absorption troughs are seen for Ly$\alpha$/N{\sc v}, Si{\sc iv}, C{\sc iv}, and Al{\sc iii}.

\section{Microlensing Analysis}
\label{sigs}

Gravitational lensing is achromatic, and so in the absence of microlensing, differential extinction, or time-dependent spectral variation, the spectra of lensed images of a multiply-imaged quasar are identical modulo a scaling factor. Microlensing is typified by a difference in the magnification of one or more lensed images compared to the time-averaged magnification for that lensed image (which we will refer to as the macrolens magnification, or just macro-magnification). %a note on why microlensing occurs
{\it Differential microlensing} is characterized by magnifications that vary between different emission regions, resulting in a change in the shape of the spectrum of the affected lensed image. Differential microlensing is sensitive to emission region size and projected location, is frequently seen between BELs and continua, and also often shows wavelength dependence within these features.

\subsection{Intrinsic Variability and Dust Extinction}

Spectral variability coupled with different light travel times can result in spectral differences between lensed images. In the case of H1413+117 the time delay is potentially significant, estimated, for example at 23.4$\pm$4 days between images A and D \citep{Goicoechea10}. This variability can potentially lead to differences in four key observables between lensed images: BAL and BEL profile shapes, continuum-to-BEL flux ratio, and the spectral index of the continuum. 
While it is difficult to extricate the differences due to intrinsic variability from those due to microlensing, we can comment on the effect of this on our analysis.

Variation in broad absorption and emission line profile shapes would be very difficult to discriminate from microlensing. Fortunately the effect is probably minimal. As we show in Section \ref{fluxratios} the BEL profiles are identical between lensed images, precluding both differential microlensing and intrinsic variability in the BELR spectrum. It is not so easy to compare BAL profiles, however we note that, while this profile has been observed to vary, this variation occurs over months and years, not days (H10). As we will see in Section \ref{microlensingbehavior} we estimate the size of the BELR (and hence also the BAL wind) at $>$100 light days, and so their intrinsic variation timescales should be significantly longer than the time delay.

Variation in the continuum-to-BEL ratio between lensed images cannot be ruled out, and we do see significant variation in this ratio, which we attribute to microlensing because it is consistent with the microlensing behavior of the lens in recent years. In Section \ref{algebra} we use this difference in continuum-to-BEL ratios to decompose the spectrum into intrinsic components. It's important to note that for the purpose of the decomposition of spectral components, it does not matter why this ratio varies between images---whether by microlensing or intrinsic variability---the decomposition is still valid as long as the BELs don't show wavelength-dependent variations. 

Variation in the slope of the continuum could also be due to intrinsic variability, and we do indeed observe a difference in slope (Sect. \ref{fluxratios}), however we don't investigate this in detail in this paper.

Differential extinction between lensed images will not affect the continuum-to-BEL ratios as our measurement of the continuum strength is made at the relevant BEL wavelength range. Nor will it significantly affect broad line profiles, which appear identical between lensed images anyway. Differential extinction only affects the continuum slope and ratios between BELs, which are not a focus of his work.

\subsection{Flux Ratios and Microlensing Wavelength Dependence}
\label{fluxratios}

\begin{figure*}[ht]
\includegraphics[width=155mm]{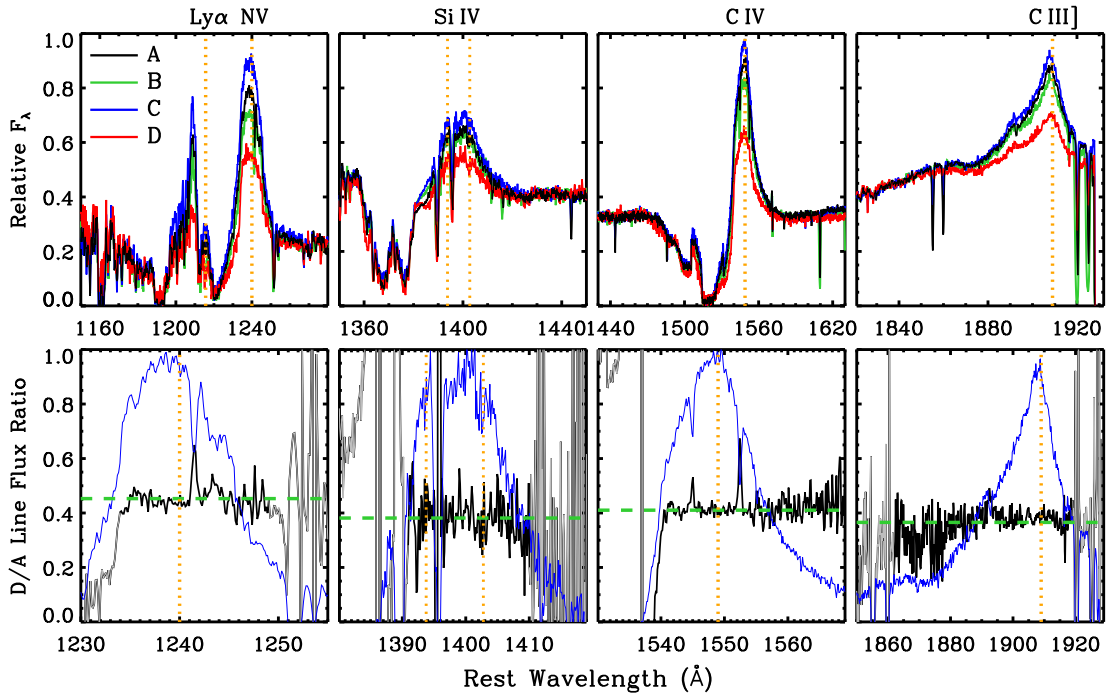}
\caption{{\it Upper}: Spectra at broad emission lines with fluxes scaled so that the continua match in regions bounding lines -- black: image A, green: image B, blue: image C, red: image D.  {\it Lower}: Ratios of the spectra of lensed images A \& D at broad emission lines after subtraction of best-fit continuum spectra (black). Blue shows the continuum-subtracted broad lines for lensed image A. Green lines in the lower panels show the average broad-line flux ratio for each line (Table \ref{fluxratiotable}). Orange lines indicate the BEL rest-frame wavelengths.
\label{linesratios}}
\end{figure*}

To inspect differential microlensing properties of H1413+117 we first obtain an estimate of the unabsorbed continuum spectrum for each lensed image by performing a linear fit to regions of the spectrum least affected by emission or absorption (Fig.~\ref{spec} shows fits for images A \& D). Prominent iron emission features force us to choose small windows at rest frame 1420---1460\AA, 1685---1692\AA, and 1795---1805\AA. While the continuum is unlikely to be linear over the full range of this spectrum, we find that the fit gives a very plausible estimate, particularly redward of the N{\sc v} line. Subtraction of the fitted continua allow us to measure the average flux ratios across each BEL for each lensed image with respect to image A. These ratios and the average flux ratios for the continuum at each line center are given in Table~\ref{fluxratiotable}. 
For the purpose of inspection of the microlensing properties, we find that changing the continuum level and slope by allowing variation of $\pm$5\% (the width of the pale gray region surrounding fits in Fig.~\ref{spec}) does not significantly affect results.

\begin{table}[t]
\begin{footnotesize}
\begin{center}
\begin{tabular}{lcccccccc}
\tableline\tableline
  &\multicolumn{2}{c}{B/A}    &&\multicolumn{2}{c}{C/A}&&\multicolumn{2}{c}{D/A}\\
                 & line & continuum & & line & continuum & & line & continuum \\
\tableline
%N{\sc v}             & 0.783  $\pm$ 0.003  & 0.92 $\pm$ 0.05 & & 0.93 $\pm$  0.02 & 0.81 $\pm$ 0.04   &   &  0.45 $\pm$ 0.01 & 0.77 $\pm$ 0.04 \\
%Si{\sc iv}            & 0.82  $\pm$ 0.01  & 0.91 $\pm$ 0.05    & &    0.930 $\pm$  0.005 & 0.79 $\pm$ 0.04    &  &    0.38 $\pm$ 0.01 & 0.75 $\pm$ 0.04 \\
%C{\sc iv}            & 0.807 $\pm$ 0.003  & 0.91 $\pm$ 0.05 & & 0.85 $\pm$  0.01 & 0.77 $\pm$ 0.04   & & 0.410 $\pm$ 0.003 & 0.73 $\pm$ 0.04 \\
%C{\sc iii]}           & 0.845  $\pm$ 0.005 & 0.88 $\pm$ 0.04    & &    0.87  $\pm$ 0.01 & 0.71 $\pm$ 0.04    & &    0.381 $\pm$ 0.002 & 0.68 $\pm$ 0.03 \\
%Fe complex        & \multicolumn{2}{c}{0.79  $\pm$ 0.01}     & &    \multicolumn{2}{c}{0.84  $\pm$ 0.01}      & &    \multicolumn{2}{c}{0.41 $\pm$ 0.01}  \\
%\tableline
%mean BEL          & \multicolumn{2}{c}{0.80  $\pm$ 0.04}     & &    \multicolumn{2}{c}{0.89  $\pm$ 0.04}     & &    \multicolumn{2}{c}{0.40 $\pm$ 0.02} \\
N{\sc v}             & 0.78  $\pm$ 0.02  & 0.92 $\pm$ 0.02   & &    0.93 $\pm$  0.03   & 0.81 $\pm$ 0.02   &  &    0.45 $\pm$ 0.03 & 0.77 $\pm$ 0.03 \\
Si{\sc iv} line     & 0.82  $\pm$ 0.02   & 0.91 $\pm$ 0.01    & &    0.93 $\pm$  0.02 & 0.79 $\pm$ 0.01    &  &    0.38 $\pm$ 0.02 & 0.75 $\pm$ 0.02 \\
C{\sc iv}            & 0.81 $\pm$ 0.01  & 0.91 $\pm$ 0.01    & &    0.85 $\pm$  0.01   & 0.77 $\pm$ 0.01   & &     0.41 $\pm$ 0.01 & 0.73 $\pm$ 0.01 \\
C{\sc iii]}           & 0.84  $\pm$ 0.02 & 0.88 $\pm$ 0.02    & &    0.87  $\pm$ 0.03   & 0.71 $\pm$ 0.02    & &    0.38 $\pm$ 0.03 & 0.68 $\pm$ 0.02 \\
Fe complex        & \multicolumn{2}{c}{0.79  $\pm$ 0.02}      & &    \multicolumn{2}{c}{0.84  $\pm$ 0.01}      & &    \multicolumn{2}{c}{0.41 $\pm$ 0.01}  \\
\tableline
mean BEL          & \multicolumn{2}{c}{0.80  $\pm$ 0.04}     & &    \multicolumn{2}{c}{0.89  $\pm$ 0.04}     & &    \multicolumn{2}{c}{0.40 $\pm$ 0.02} \\
%\tableline
mean cont.&  \multicolumn{2}{c}{0.91 $\pm$ 0.03}  & & \multicolumn{2}{c}{0.77 $\pm$ 0.02} & & \multicolumn{2}{c}{0.73 $\pm$  0.02} \\
\tableline\tableline
\end{tabular}
\caption{Measured flux ratios for broad emission lines and continuum at BEL locations for lensed images B, C, and D with respect to A, based on linear fit to continuum (see Sect. \ref{fluxratios}), and avoiding absorption features. Also included are mean BEL and continuum flux ratios.
\label{fluxratiotable}}
\end{center}
\end{footnotesize}
\end{table}

Figure~\ref{linesratios} (upper panels) compares the spectra of all lensed components around the broad emission lines. To highlight the effect of differential microlensing we scale the spectra of lensed images B, C, \& D so that the average continuum emission at each line (from the continuum fits) matches that of image A.
We see significant differential microlensing between broad emission lines and continuum in lensed image D versus all other images, with the emission lines at lower magnification in D relative to its continuum. Other lensed images show weak differential magnification between lines and continuum. Due to an unknown lens redshift the lens model is not well established and so the macro-magnifications are not well known. For this reason we don't know which, if any of A, B, or C are free of microlensing. Nonetheless, we can conclude that image D is undergoing significant differential microlensing between BELR and accretion disk.

These results are consistent with observations over the past decade (Angonin  et al. 1990, {\O}stensen  et al. 1997, Monier  et al. 1998, Anguita  et al. 2008, H10), which show strong anomalous line-continuum flux ratios in image D compared to A, B, and C.

Our observations show wavelength-dependent differential microlensing of the continuum in the D/A and C/A ratios. The D/A ratio tends to more extreme values (deviating from the average BEL ratio) towards shorter continuum wavelength, as was also observed by H10. It is tempting to interpret this as a simple inverse correlation between emission region size and microlensing intensity, however we note that the observed C/A ratios have the opposite behaviour, tending towards the average C/A BEL ratio towards {\it longer} wavelengths. In Section \ref{microlensingbehavior} we show that we cannot expect a monatonic trend of magnification ratio with size across multiple orders of magnitude in emission region size. Proper analysis of differential microlensing across the continuum requires careful lensing simulation, however we do not perform that analysis in this work.

To investigate the microlensing state of the BELs we determine the wavelength-dependent flux ratios for the BEL SEDs for lensed images A and D. These are just the ratios of the spectra of images A and D after subtracting the continuum estimates described above. In Figure~\ref{linesratios} (lower panels) we can see that the flux ratios for the N{\sc v}, Si{\sc iv}, C{\sc iv}, and C{\sc iii]} broad emission lines are flat over the wavelength range redward of BAL absorption. Ly$\alpha$ is omitted due to its overlapping the N{\sc v} absorption trough, which makes continuum subtraction impossible. This flatness indicates that there is no differential microlensing across the BELR velocity structure redward of the BAL absorption, and very little difference between these emission species. The mean and individual BEL flux ratios are given in Table~\ref{fluxratiotable}.  Error intervals include standard error of the mean and uncertainty in the continuum fits. Variation of the level of continuum subtraction up to its estimated uncertainty at the location of each emission line does not change the observed flatness of these flux ratios.

\subsection{Comparison to Q2237+0305 and Size Constaints on the BELR}
\label{microlensingbehavior}

In \citet{ODowd11} we analyzed microlensing signatures in the C{\sc iv} and C{\sc iii]} BELs of Q2237+0305, the Einstein Cross. In that case there was clear differential microlensing across the velocity structure of these broad lines, with magnification ratios varying by $\sim$25-30\% between line center and wings. The BEL magnification ratio approached the highly anomalous continuum magnification ratio at the highest velocities. Our simple model indicated that this trend was consistent with the broad line flow being a Keplerian field, and not supporting a pure radiation-driven outflow without a substantial gravitational component. This differential microlensing signature sharply contrasts the wavelength-independent BEL magnification ratios seen in H1413+117 (Sect.~\ref{sigs}). This wavelength independence suggests that the BELR in H1413+117 is large enough to average over fluctuations in the magnification surface. Given the very low redshift of the lensing galaxy in Q2237+0305 ($z_L=0.04$) and its commensurately large fluctuation scale-length, differential microlensing of its BELR is not surprising.
To explore the implications of this difference in more detail we compare the expected microlensing behavior of these two quasars as a function of source size.

Although the lensing galaxy of H1413+117 does not have a confirmed redshift, it is possible to explore plausible microlensing scenarios using estimates from literature: z$_L$ = 0.94 based on the probability distribution from lens parameters \citep{Mosquera11} and supported by the z$_L\sim$ 0.9 photometric redshift of overdensities in the field \citep{Kneib98, Faure04}; and z$_L$ = 1.4 \citep{Surdej93, Witt95}, that is also comparable to the z $\sim$ 1.5 absorption systems in the quasar spectrum \citep{Magain88, Angonin90}.
For the H1413+117 lens macro-model we use the convergence and shear estimates of \citet{Witt95} ($\beta = 1$ case), which are nearly identical to those of \citet{Mediavilla09}. We assume an 80\% smooth matter percentage, consistent with the values estimated in other lenses where the line of sight does not pass through the galactic core \citep{Keeton06, Chartas09, Pooley09, Dai09, Bate11}. For Q2237+0305 we use the macro-model of \citet{Trott10}, and Trott \& Wayth (priv. comm.). In this case lensed images are clustered around the star-dominated galactic core and microlensing is consistent with a 0\% smooth matter percentage \citep{Bate11}, which we employ here.
We obtained simulated maps of microlensing magnifications from the GERLUMPH database \citep{gerlumph} based on these models.

We simulate the expected microlensing behavior of these lenses, using Monte Carlo simulations. Emission region surface brightness distributions are simulated with Gaussian profiles whose half-light radii represent emission region sizes. Note that except in the case of very strong microlensing events, half-light radius rather than the detailed surface brightness distribution governs microlensing properties \citep{Mortonson2005}. 
A pair of random locations on Gaussian-convolved magnification maps (one map for each of a pair of lensed images) yields a possible magnification ratio between those lensed images. By varying half-light radius and location we can then explore microlensing behavior as a function of emission region size. We determine D/A ratios for H1413+117 at 50,000 location pairs and B/A ratios for Q2237+0305 at 25,000 location pairs. We explore sizes from $3\times10^{13}$~m to $2\times10^{15}$~m, constrained by the resolution and size of the GERLUMPH maps. 

\begin{figure*}%[ht]
\includegraphics*[width=90mm]{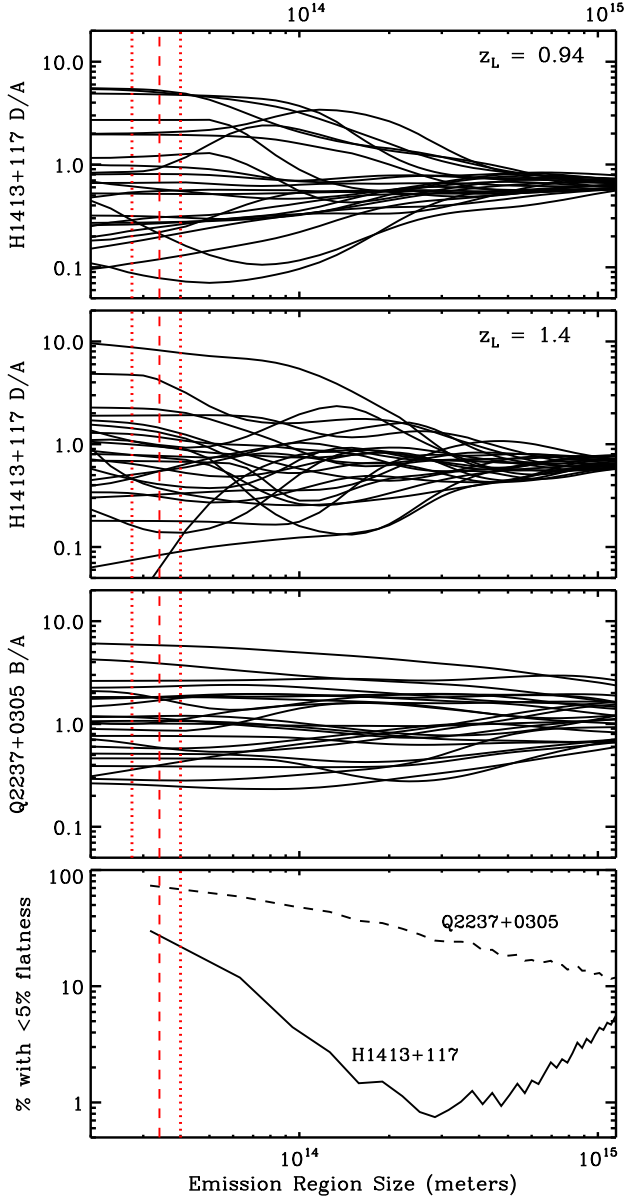}
\caption{{\it Top panel}: Simulated magnification ratios as a function of source size for H1413+117, lensed images A and D, with an assumed lens redshift $z_L= 0.94$ (showing 26 of 50,000 simulated tracks). {\it 2$^{nd}$ panel}: As above, but for $z_L=  1.4$.  {\it 3$^{rd}$ panel}:  As above, but for Q2237+0305 ($z_L=  0.04$; showing 26 of 25,000 simulated tracks). 
{\it Bottom panel}: Percentage of simulated tracks that are flat as a function of maximum emission region size. Flatness is defined as $< 5$\% variation in magnification ratio over 0.5 dex in size up to the maximum emission region size.
The red dotted line shows the half-light radius of a Shakura-Sunyaev thin disk accreting at the Eddington limit around a non-rotating black hole with mass $4\times10^8$ M$_{\odot}$ (left line) and $10^9$ M$_{\odot}$ (right line), observed at 150~nm. The red dashed line is the 150~nm continuum sourse size estimate of H10.
\label{ratiovssize}}
\end{figure*}

Figure~\ref{ratiovssize} (top 3 panels) shows magnification ratio versus emission region size for a random selection of 26 of these locations.  Magnification ratios are normalized to unity at the macro-model magnification ratios (the large-scale average ratios).
These figures, which we will refer to as {\it convergence plots}, illustrate two important points: 

1) As expected, magnification ratios will converge on the macro-model ratios above a certain emission region size (the {\it convergence size}).
We see that for H1413+117 (top 2 panels) the D/A convergence size is $\sim 5\times10^{14}$~m ($\sim$20 light days) for both the $z_L = 0.94$ and $z_L = 1.4$ cases. If the wavelength independence that we observe in the BEL magnification ratios (Sect. \ref{fluxratios}) is due to such convergence, then this gives us an approximate lower limit on the size of observed BELRs. 
In the case of Q2237+0305 (panel 3) the convergence of the B/A magnification ratio does not occur until significantly greater than $10^{15}$~m, beyond the limit of our simulated magnification maps, and so the observation of differential microlensing across the full velocity width of its BELR is not surprising.

2) We expect tracks on the convergence plot to be flat for sizes larger than the convergence size, however we also see many flat tracks for very small emission regions. This independence of magnification ratio with size often occurs in cases where the magnification ratio deviates significantly from the macro-model value (i.e. where there is strong microlensing). For H1413+117, most simulated tracks are relatively flat below $\sim 10^{14}$~m. There is an intermediate range over which all tracks converge on the macro-model ratios---the {\it convergence range}. The convergence range spans $\sim 10^{14}$~m to $\sim 5\times10^{14}$~m for H1413+117, and in it we see strong size dependence of magnification ratios, and here the only flat tracks are those already close to the macro-model ratio. 

The interpretation of this behavior is straightforward: The small-scale flatness seen the convergence plots occurs when a small emission region falls in relatively featureless regions between strong magnification gradients in the caustic network. As size increases into the convergence range, our larger source is beginning to overlap high or low magnification structures in the caustic network, resulting in size-dependent, differential microlensing. When the emission region is large enough to sample many such structures, tracks again flatten at the macro-model magnification ratio.

The magnification ratios of the C{\sc iv} and C{\sc iii]} BELs in H1413+117 vary by less than 5\% across the unabsorbed component of their velocity structure (Sect. \ref{fluxratios}), and so must reside on a ``flat'' region of a convergence plot track. We can place more formal constraints on the possible sizes of these emission regions by determining the number of tracks that show similar flatness over a given size range. We conservatively assume that the broad line gas flow spans a size range of 0.5 dex across its entire velocity structure. In fact the BELR probably spans a larger range, but choosing a smaller range yields more tracks within our flatness constraint, and so produces weaker (more conservative) size constraints. 

Figure~\ref{ratiovssize} (bottom) shows the percentage of tracks consistent with $<$5\% variation in the magnification ratio over 0.5 dex in emission region size as a function of the maximum size of the emission region, for both H1413+117 ($z_L = 0.94$) and Q2237+0305. 

In the case of Q2237+0305 Figure \ref{ratiovssize} shows a reasonable likelihood of observing size-independent microlensing over this full range of emission region sizes. For H1413+117, size-independent microlensing is unlikely ($\lesssim$ 5\%) for BELRs spanning 0.5 dex with maximum size in the range $\sim 10^{14}$~m to $\sim 10^{15}$~m ($\lesssim$ 1\% for $\sim 2\times10^{14}$~m to $5\times10^{14}$~m). At larger sizes, size independence will become more probable, and extrapolating our results we estimate that size independence will become likely ($>$50\%) at BELR sizes of 2-3$\times10^{15}$~m, or around 100 light days. It is important to note that H10 also observe this size-independent microlensing of the BELR at a different epoch, greatly decreasing the probability that the BELR is in the $\sim 10^{14}$~m to $\sim 10^{15}$~m range. For a reasonable probability of observing size independent magnification ratios in the BELR of H1413+117 in two independent epochs, the BELR probably needs to be larger than $10^{15}$~m ($\sim$40 light days), and any significantly luminous component likely arises well above the convergence size of $\sim 5\times10^{14}$~m ($\sim$20 light days). These constraints are for the $z_L = 0.94$ case; larger values of $z_L$ push us to larger BELR sizes.

We can also consider the possibility that the entire BELR is emitting at sizes smaller than $\sim 10^{14}$~m ($\sim$4 light days). We think this unlikely for two related reasons: 1) our observations and those of H10 show  wavelength-dependent differential microlensing in the continuum D/A ratio, which precludes flat tracks at small size needed for size independence in the BEL magnification ratios; 2) for a high likelihood of size independence at small scales at two epochs, the emission region must be closer to a few times $10^{13}$~m, which is approaching the scale of the accretion disk (see Sect.~\ref{sizes}), and yet the D/A ratios indicate very strong differential microlensing between continuum and BELs at multiple epochs (see Sect. \ref{fluxratios}). 

From the reverberation mapping BELR size-luminosity relation for luminous quasars of \citet{Kaspi07}, H1413+117 (with its $\lambda$~F$_\lambda$(1350\AA)$ = 4\times10^{45}$~erg~s$^{-1}$) is expected to have a C{\sc iv} BELR in the range of 25--90 light days. From their microlensing study of of 16 lensed quasars (not including H1413+117), \citet{Guerras13} find a BELR size range for high-ionization lines in their luminous subsample of $36^{+30}_{-14}$ light days. Both of these results support our estimate of a BELR that is signficantly larger than 20 light days, and very unlikely to be $\lesssim$4 light days.

\subsection{Black Hole Mass and Accretion Disk Size}
\label{sizes}

We estimate the black hole mass in H1413+117 at $4\times10^8$~M$_\odot$ based on the C{\sc iv} width and using the \citet{Vestergaard06} M$_{BH}$--FWHM relation. We measure the C{\sc iv} FWHM of 3,400~km/s based on the unabsorbed, red side. Note that the intrinsic C{\sc iv} BEL is either symmetric or shows a blue excess, as discussed in Section \ref{belrcomp}, and so the correct FWHM and corresponding black hole mass may be a larger.

In order to obtain an upper limit estimate on the size of the accretion disk in H1413+117 we assume a Shakura-Sunyaev thin disk \citep{SS} with a $10^9$~M$_\odot$ non-rotating black hole accreting at the Eddington limit, and observed at 150~nm. This yields a source half-light radius of $4\times10^{13}$~m, shown in Figure~\ref{ratiovssize}. For our black hole mass estimate of $4\times10^8$~M$_\odot$ and the same accretion parameters the 150~nm source size is $2.7\times10^{13}$~m. Based on differential microlensing across the continuum, H10 estimate a continuum source size of $3.4\times10^{13}$~m.

The accretion disk for both quasars is much smaller than the convergence range. For H1413+113 both size-dependent and size-independent magnification ratios are reasonably common at small radii, and it's not surprising that the continuum shows some differential magnification in both our observations and those of H10. This is also true for Q2237+0305, however size dependence occurs less frequently at these small size scales, in $\sim$20\% of cases. 

The D/A magnification ratio in the H1413+117 continuum tends towards the average D/A BEL ratio towards longer continuum wavelengths (although never getting close to the BEL ratio), and so it's tempting to interpret this as the accretion disk becoming large enough to start averaging over causic structures and hence approach the macro-model magnifications. However Figure \ref{ratiovssize} shows that even a large accretion disk is much smaller than the convergence range. In this regime, magnfication ratio can have a size dependence that either approaches the macro-model value or moves away from the macro-model value, even if all tracks ultimately converge on the same value. In short, we cannot interpret a trend towards the macro-model magnifications as a reliable proxy for increasing emission region size {\it except} in the convergence range. It is therefore not surprising that the C/A continuum ratios tend in the opposite direction to the D/A ratios, with shorter wavelengths approaching the C/A BEL ratio.

Proper interpretation of the differential microlensing in the continuum requires higher resolution magnification maps and a careful statistical analysis, which we leave to a later study.

\section{Algebraic Decomposition of Intrinsic Spectral Components}
\label{algebra}

\subsection{Decomposition of Components}

\begin{figure*}[ht]
\includegraphics*[width=155mm]{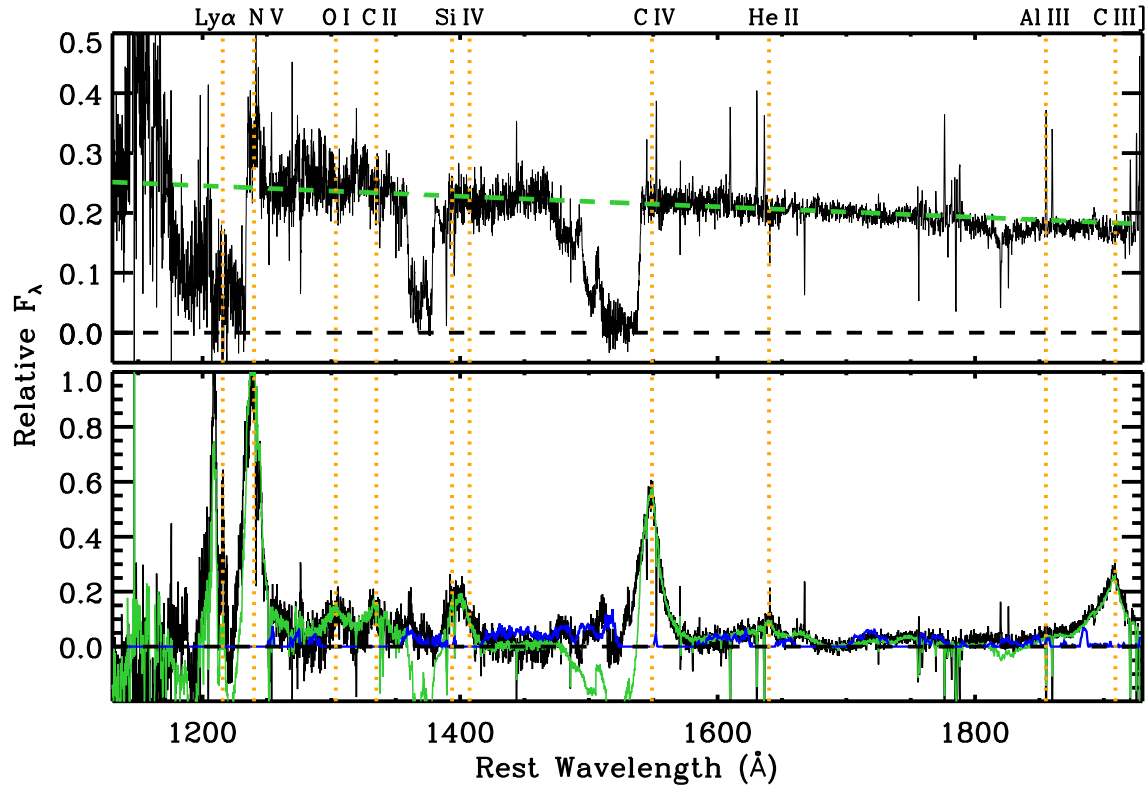}
\caption{{\it Upper}: Absorbed continuum spectrum of lensed image D in H1413+117 ($\boldsymbol{\mathcal C}_D$ from Eqn. \ref{cont}; black). Green dashed line shows the continuum fit to image D. {\it Lower}:  absorbed broad emission line spectrum of lensed image D ($\boldsymbol{\mathcal L}_D$ from Eqn. \ref{line}; black). The green line shows the observed spectrum of image D with continuum fit subtracted. The blue line shows the iron emission template of \citet{VW01} with arbitrary scaling.
\label{decomp}}
\end{figure*}
 
We can consider the spectra of the lensed images in H1413+117 to be linear combinations of intrinsic spectral components from different emission regions, each subject to absorption and lensing that may vary between lensed images and emission regions. If the relative strengths of the emission components vary between one lensed image and another due to differential microlensing or intrinsic variability, then with some assumptions it can be possible to solve for these components. 

\citet{Angonin90} first apply this approach to H1413+117 to derive an intrinsic BAL absorption profile for Si{\sc iv}, C{\sc iv},and Al{\sc iii}. They describe the spectrum as composed of physical components: a continuum + BEL.
H10 explore the approach in more detail, applying the method of \citet{Sluse07} to H1413+117 to extract Ly$\alpha$+N{\sc v}, Si{\sc iv}, C{\sc iv} BAL troughs. In the $F_{\mu}$ method, the quasar spectrum is explicitly described in terms of a component that is only macrolensed and one that is both micro- and macrolensed.
Both approaches lead to similar results, and this is because their assumptions are similar: that the continuum is microlensed but the BELR is not; as a result the micro+macrolensed component of H10 is essentially the absorbed, lensed continuum.

Here we derive a new spectral decomposition of H1413+117. Our approach is mathematically similar to both Angonin  et al. and H10, and we discuss their connection in \ref{H10comparison}. We choose to formulate the decomposition in terms of physical emission regions rather than by their degree of microlensing. We do this because: 1) our observation of almost constant magnification ratios across the emission line velocity structures (see Sect.~\ref{sigs}) ensures a straightforward extraction of the BEL component, and 2) we include in our formulation a separate BAL absorption term for both the continuum and BELR, which we feel adds clarity to the interpretation of the derived components, and 3) any source of variation---microlensing or intrinsic variability coupled with time delay---enables a decomposition, and yet the possible presence of the latter means that results can not be easily interpreted as microlensed/non-microlensed components. 

We describe the spectrum \textbf{\textit{F}}$_X$ of lensed image X as a linear combination of two components: 1) $\boldsymbol{\mathcal C}_{X}$: the intrinsic continuum spectrum \textbf{\textit{F}}$_C$ multiplied by the BAL absorption profile affecting the continuum \textbf{\textit{A}}$_C$ and the (potentially wavelength-dependent) magnification \textbf{\textit{m}}$_{C,X}$ affecting the continuum in image X, and 2) $\boldsymbol{\mathcal L}_{X}$: a broad emission line spectrum \textbf{\textit{F}}$_L$ with its own, potentially different absorption profile \textbf{\textit{A}}$_L$ and lensing magnification \textbf{\textit{m}}$_{L,X}$. 
The \textbf{\textit{F}} and \textbf{\textit{A}} components are intrinsic to the quasar and identical for all lensed images (modulo quasar variability, but see below). The lines of sight to different lensed images do not diverge significantly within the BAL wind (see Sect. \ref{lineofsight}) and so all lensed images see the same absorption from this outflow. However separate absorption terms are required becaues the accretion disk and BELR are viewed through different regions of the BAL wind. The magnification factor is different for each lensed image and in the presence of microlensing this factor may differ between the continuum and BELs, and each of these may have wavelength dependence. 
We have neglected to include an explicit term for reddening or to account for intrinsic variability in the quasar. Both of these effects cause wavelength-dependent multiplicative changes to component spectra, and so in terms of the decomposition we can consider these to be folded into the magnification terms; the origin of these multiplicative effects does not affect the decomposition. For lensed images 1 and 2 we have:

\begin{equation}
{\bf F}_1 = {\bf m}_{C,1} {\bf A}_C {\bf F}_C + {\bf m}_{L,1} {\bf A}_L {\bf F}_L\\
\end{equation}

\begin{equation}
 {\bf F}_2 = {\bf m}_{C,2} {\bf A}_C {\bf F}_C + {\bf m}_{L,2} {\bf A}_L {\bf F}_L
\end{equation}

The only known quantities in these equations are the observed spectra, \textbf{\textit{F}}$_1$ and \textbf{\textit{F}}$_2$. Although we don't know the intrinsic magnification factors of each component, it is possible to measure their ratios. In the absence of differential absorption or reddening these are just the wavelength-dependent flux ratios for each component. We set ${\boldsymbol\eta}_C = {\bf m}_{C,1}/{\bf m}_{C,2}$ and ${\boldsymbol\eta}_L$ = \textbf{\textit{m}}$_{L,1}/$\textbf{\textit{F}}$_{L,2}$: the magnification ratios in continuum and emission line. We can then solve for $\boldsymbol{\mathcal C}$ and $\boldsymbol{\mathcal L}$ for one lensed image:

\begin{equation}
\label{cont}
\boldsymbol{\mathcal C}_{1} = {\bf m}_{C,1} {\bf A}_C {\bf F}_C = \frac{{\boldsymbol\eta}_C}{{\boldsymbol\eta}_C - {\boldsymbol\eta}_L} ({\bf F}_1 - {\boldsymbol\eta}_L {\bf F}_2)
\end{equation}

\begin{equation}
\label{line}
\boldsymbol{\mathcal L}_{1} = {\bf m}_{L,1} {\bf A}_L {\bf F}_L = \frac{{\boldsymbol\eta}_L}{{\boldsymbol\eta}_L - {\boldsymbol\eta}_C} ({\bf F}_1 - {\boldsymbol\eta}_C {\bf F}_2)
\end{equation}

We can think of these equations as representing a scaled subtraction of the spectrum of one lensed image (\textbf{\textit{F}}$_2$) from another (\textbf{\textit{F}}$_1$), with the scaling factor ${\boldsymbol\eta}$ giving complete subtraction of one spectral component. In Eq. \ref{cont}, for example, \textbf{\textit{F}}$_2$ is scaled so that its emission line flux matches that of \textbf{\textit{F}}$_1$. As long as the line-to-continuum ratio is different between \textbf{\textit{F}}$_1$ and \textbf{\textit{F}}$_2$ then \textbf{\textit{F}}$_1 - {\boldsymbol\eta}_L$~\textbf{\textit{F}}$_2$ eliminates the BEL spectrum leaving a representation of the absorbed continuum spectrum with a slope somewhere between that of lensed images 1 and 2. The scaling factor ${\boldsymbol\eta}_C/({\boldsymbol\eta}_C - {\boldsymbol\eta}_L)$ then recovers the normalization and slope of the continuum spectrum seen in image 1. A similar interpretation of Eq. \ref{line} eliminates the continuum component to give the lensed, absorbed BEL spectrum for image 1.

For our decomposition of H1413+117 we define \textbf{\textit{F}}$_1$ to be the spectrum of image D and \textbf{\textit{F}}$_2$ to be that of image A.
We take ${\boldsymbol\eta}_C$ to be the wavelength-dependent continuum ratio from the fits described in Section~\ref{sigs}. As seen in Sect.\ref{sigs}, the BEL magnification ratios are not wavelength dependent over each line and vary relatively little between Si{\sc iv}, C{\sc iv}, and C{\sc iii]}. To reconstruct spectral components over the full wavelength range of the spectra we take ${\bf \eta_L}$ to be a constant: $\eta_L = 0.40 \pm 0.02$, the mean continuum-subtracted broad line flux ratio from Table~\ref{fluxratiotable}. For individual BEL decompositions we use their individual magnification ratios (see below).

Figure~\ref{decomp} shows the resulting decomposition over the full spectrum. 
The upper panel shows $\boldsymbol{\mathcal C}_{D}$: the lensed, absorbed quasar continuum (Eq.~\ref{cont}), with the continuum magnification factor of image D. We overplot the continuum fit for the image D spectrum.
The lower panel shows $\boldsymbol{\mathcal L}_{D}$: the lensed, absorbed BEL spectrum (Eq.~\ref{line}), again as magnified in image D. We overplot the observed spectrum of image D minus the continuum fit and also an iron emission spectrum from the templates of \citet{VW01}, scaled to roughly match the iron features in the spectrum.

\begin{figure*}[ht]
\includegraphics*[width=140mm]{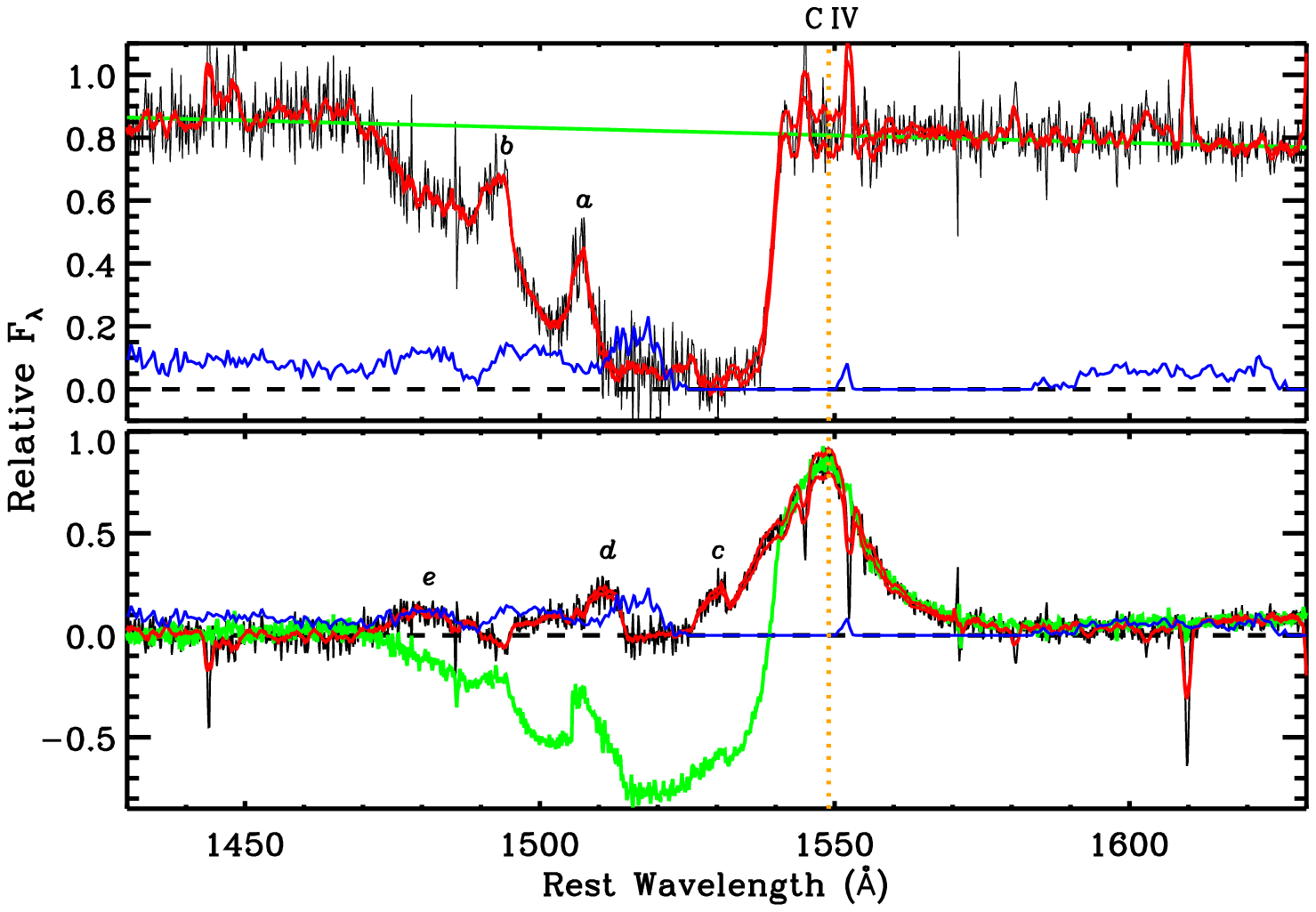}
\caption{{\it Upper}: Absorbed continuum spectrum of lensed image D in H1413+117 around C{\sc iv} ($\boldsymbol{\mathcal C}_D$ from Eqn. \ref{cont}; black). Green dashed line shows the continuum fit to image D. {\it Lower}:  absorbed C{\sc iv} BEL spectrum of lensed image D ($\boldsymbol{\mathcal L}_D$ from Eqn. \ref{line}; black). The green line shows the observed spectrum of image D with continuum fit subtracted. The blue line shows the iron emission template of \citet{VW01} with arbitrary scaling.
 Red lines in both panels show the upper and lower limits of range of possible decompositions based on the uncertainties in the continuum fit and measured flux ratios. The blue line shows the iron emission template of \citet{VW01}. Letters are designations for absorption trough features discussed in Section \ref{features}.
\label{civdecomp}}
\end{figure*}

Figure~\ref{civdecomp} shows a nearly identical decomposition as Fig.~\ref{decomp} over the range of the C{\sc iv} line, with the only difference being that we take $\eta_L = 0.410 \pm 0.003$, the continuum-subtracted flux ratio over the red wing of the C{\sc iv} line. Fig.~\ref{civdecomp} also shows the range of uncertainty for these decompositions (red lines): the maximum and minimum decomposed BEL and continuum components obtained by allowing the continuum flux to vary by $\pm 5$\% (see Sect.~\ref{sigs}) and $\eta_L$ to vary over its uncertainty range. It can be seen that the general shape of the decomposed components does not change significantly over the range of possible decompositions.

\begin{figure*}[ht]
\includegraphics*[width=140mm]{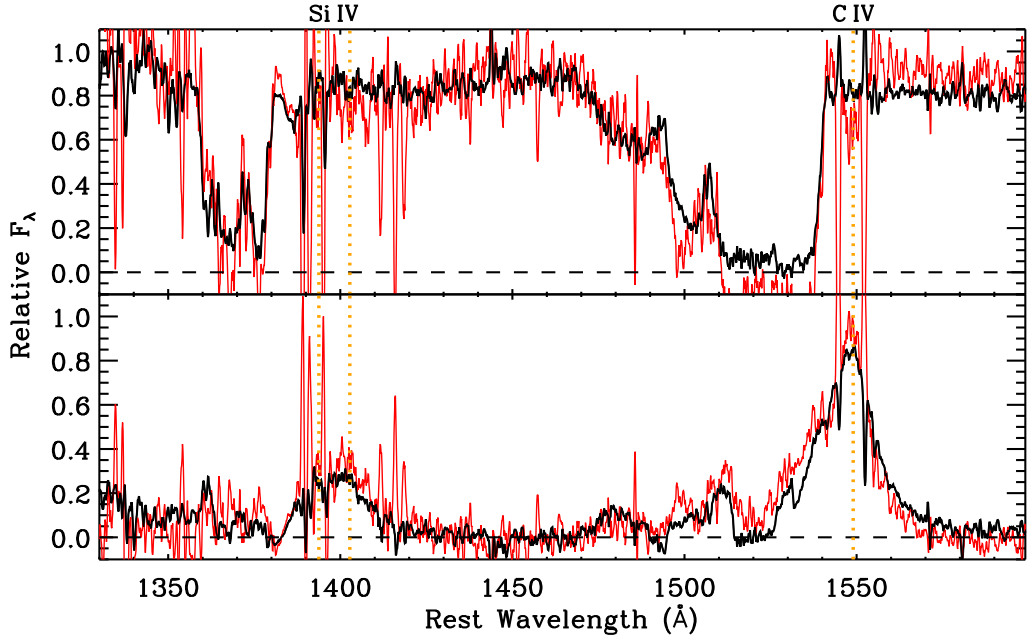}
\caption{Decomposed Si{\sc iv} to C{\sc iv} BEL ({\it upper}) and continuum ({\it lower}) spectral components of H1413+117 derived from lensed images A and D ({\it black}) and from lensed images B and C ({\it red}).
\label{comparedecomp}}
\end{figure*}

Lens image D is experiencing by far the strongest microlensing between the continuum source and BELR. However we do see some differences in broad line to continuum flux ratios in A, B and C. 
Figure~\ref{comparedecomp} shows these same decompositions over the Si{\sc iv} and C{\sc iv} lines for image pairs A \& D and for B \& C. In these cases we take the A/D and B/C emission line flux ratios to be the average for the Si{\sc iv} and C{\sc iv} lines for each image pair. We see very consistent results, albeit with significantly more noise in the B \& C decomposition due to the smaller differences between these spectra.

\begin{figure}[h]
\includegraphics*[width=120mm]{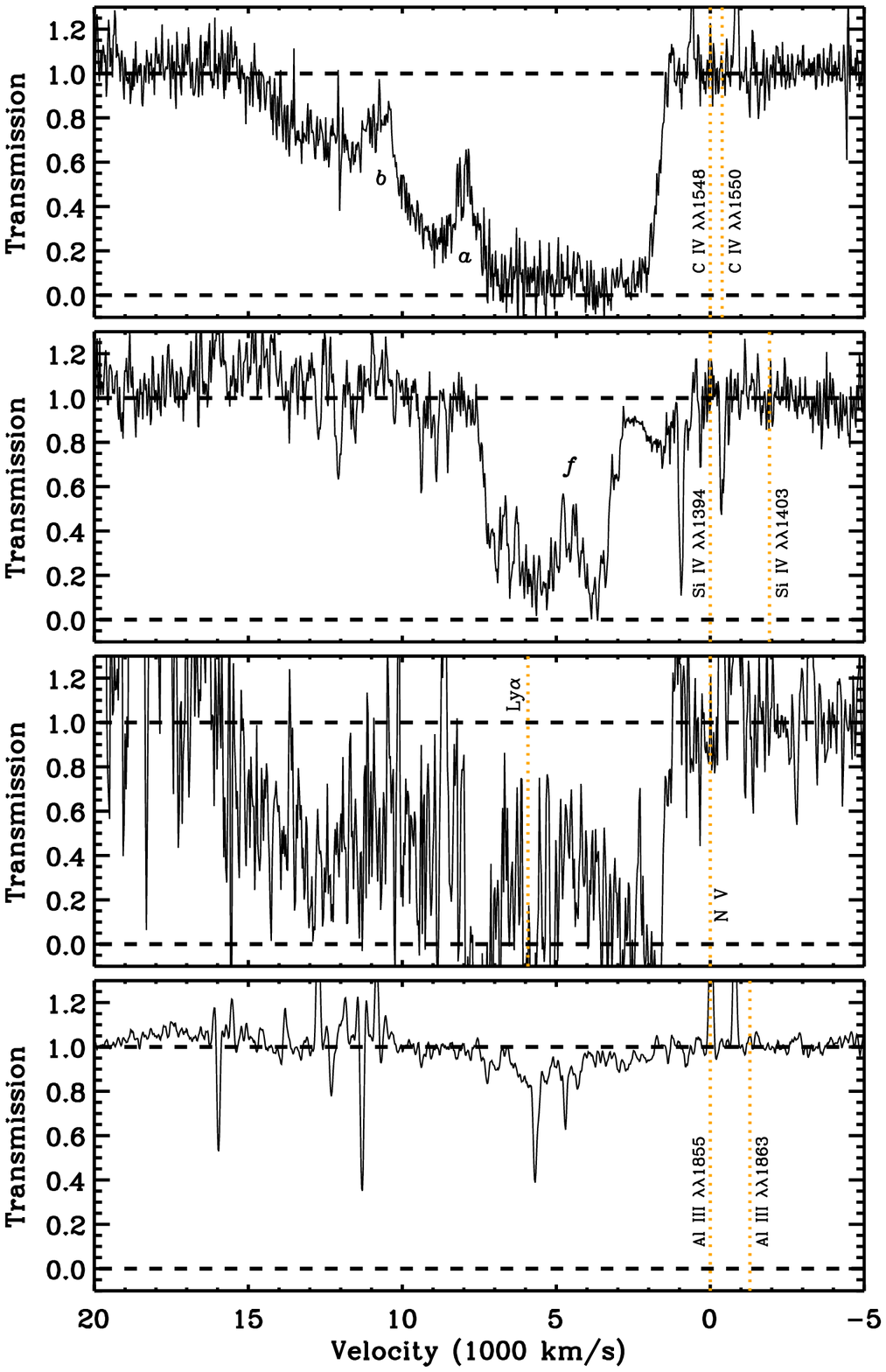}
\caption{Intrinsic absorption spectrum of BAL wind for (from top to bottom) C{\sc iv}, Si{\sc iv}, N{\sc v}, and Al{\sc iii} based on continuum spectrum derived from Eqn. \ref{cont}. Velocity zero-point is set at line center or at lower wavelength line center for doublets.
Vertical dotted lines show line centers. Letters are designations for absorption trough features discussed in Section \ref{features}.
\label{pcygni}}
\end{figure}

If we divide $\boldsymbol{\mathcal C}_D$, the lensed, {\it absorbed} continuum spectrum, by our estimate of the lensed, unabsorbed continuum \textbf{\textit{m}}$_{C,D}$~\textbf{\textit{F}}$_C$  we recover an estimate of the intrinsic absorption profile affecting the continuum. Figure \ref{pcygni} shows this for the C{\sc iv}, Si{\sc iv},  N{\sc v}, and Al{\sc iii} absorption troughs. 

\subsection{Comparison of Decomposition Approaches}
\label{H10comparison}

The derived spectral components are similar to those found by H10, and the derived absorption profile is similar to that of \citet{Angonin90}.  We discuss briefly similarities and differences in these works.

H10 decompose in terms of macrolensed-only (their $F_{M}$) and microlensed + macrolensed ($F_{M\mu}$) components, and use the combined spectra of lensed images A and B ($F_{AB}$) as well as that of image D ($F_D$).
The spectral subtraction scaling factors used by H10 are $M$, the magnification ratio of the macrolensed-only component, and $A$, the  magnification ratio of the microlensed + macrolensed component. $A$ is taken as the continuum ratio in regions unaffected by broad line flux, and so is similar to our ${\boldsymbol\eta}_C$. They measure a single value of $A$ for each emission line from the average continuum bounding that line. This accounts for changes in the continuum magnification ratio from one line to the next, but not for variation across the wavelength range of the lines, however the latter is of order 1-2\% and so is not a large effect.
$M$ is determined under the assumption that at least part of each broad emission line is not microlensed, so that within the wavelength range of each line there is a region of the reconstructed micro+macrolensed component ($F_{M\mu}$) that is equivalent to the continuum spectrum. This can be interpreted as subtraction of $F_{AB}$ from $F_D$, with  $F_{AB}$ scaled to eliminate at least part of the emission line (the part of the line closest to the continuum magnification). In practice, because the emission line magnifications are constant across their wavelength range (see Sect. \ref{sigs}), this should mean almost complete subtraction of emission lines and produces a result that is very similar to our Equation 2.

\citet{Angonin90} (A90) describe their decomposition in less detail than H10, however their approach is mathematically similar to both ours and that of H10. They express the spectrum of lensed image D to be that of the (purportedly) unmicrolensed A+B+C composite plus a microlensed, absorbed continuum component. After deriving a continuum spectrum from the A+B+C composite they fit parameters to yield the absorption profile. This appears to be equivalent to performing a scaled subtraction of A+B+C from D to eliminate the BEL component and at the same time scaling out the wavelength-dependent continuum, reddening, lensing magnifications.

In summary, although A90 and H10 formulate decompositions differently and make some different assumptions, all three approaches are mathematically similar and yield similar decompositions. 
It is worth noting that in our formulation the interpretation of the decomposition does not depend on their being any macrolensed-only component, only that the continuum-BELR ratio differs between the two lensed images. While both H10 and A90 assume that the BELR is not microlensed, any such wavelength-independent microlensing or variability-related differences in the BELR do not invalidate the decompositions.

We compare the details of these decomposition results in Section~\ref{disc}.

\section{Interpretation of Decomposition Components}
\label{disc}

\subsection{Interpretation of Broad Absorption Characteristics}

The continuum spectral component (Fig.~\ref{decomp} \& \ref{civdecomp}, upper panels) and the absorption profile that we derive from it (Fig.~\ref{pcygni}) provide a picture of the BAL absorption along the line-of-sight column to the accretion disk, and 
the shape of these profiles allows us to draw specific conclusions about the structure and kinematics of the outflow. We focus on interpretations in the context of rotating disk-wind models, in which the material responsible for both BAL and BEL features arises from the accretion disk and is driven out by radiation and possibly gas pressure \citep{Murry95, Proga00, Elvis00}. %BEL or 
In the case of C{\sc iv} the continuum is experiencing essentially no absorption until $\sim$1,500~km/s blueward of the emission line central wavelength, at which point transmission drops sharply over $\sim$600~km/s to opacity $\sim 1$. This near-black region extends to $\sim$7,500~km/s before transmission ramps roughly linearly to the end of the absorption feature at $\sim$15,000~km/s blueward of line center. Two narrow emission peaks are seen in this latter region. In the case of Si{\sc iv}, absorption also has a sharp onset blueward of the Si{\sc iv} $\lambda\lambda$1394 line center, in this case at $\sim$3000~km/s. The deep region of its absorption trough ends much more sharply than in the case of C{\sc iv} at $\sim$7,500~km/s. A single clear emission peak is observed in the deep part of the trough. A weaker absorption trough appears redward of the deeper trough onset and may result from Si{\sc iv} $\lambda\lambda$1403. In any case, all absorption is blueward of the Si{\sc iv}  doublet. As with C{\sc iv}, the N{\sc v} trough begins at a blueshifted $\sim$1,500~km/s, while the Al{\sc iii} trough appears at $\sim$4,500-5,000~km/s, though may show some absorption from $\sim$2,000~km/s.

Of particular interest is the sharp, blueshifted onset of this absorption, which is especially striking in C{\sc iv}. This sharpness is not ambiguous, and is apparent over the full range of uncertainty in the decomposition (Fig.~\ref{civdecomp}, lower). While BALs with highly blueshifted absorption onsets are sometimes observed (so-called detached troughs, blueshifted at several thousand km/s; e.g. Turnshek 1987), our decomposition of H1413+117 provides the first unambiguous description of a sudden, blueshifted absorption onset that is still within the velocity range of the BEL. The steepness and blueshift of this onset tells us two things: 1) the outflow has already accelerated to some degree before either it enters our sightline or it enters the region of efficient absorption within our sightline; and 2) the kinematics of absorbing gas in our sightline is strongly dominated by outflow, not orbital motion. 

\citet{Murry95b} intepret highly detached absorption troughs as resulting from sightlines that graze the upper layer of an equitorial outflow, hence missing the vertical base of their wind model where the flow has a small radial component. 
\citet{Higginbottom13} (H13) apply radiative transfer code to their disk-wind model to predict C{\sc iv} BAL profiles, also finding that these same grazing sightlines produce a very sharp, blueshifted onset. In these equitorial disk-winds we expect any deeper line of sight---those passing near the launch region of the wind---to show absorption at low velocities. Not only do they encounter launching gas prior to significant outward acceleration, but they also encounter gas with very high rotational velocities in the vicinity of the accretion disk. This latter effect is important, as shown by H13; at launch, the BAL flow shares the Keplerian velocity of the accretion disk and so is dominated by rotational motion. If the inner disk is occluded by this gas along our line of sight, and if this gas is significantly absorbing in the vicinity of the accretion disk, then some fraction of the continuum will be absorbed by the component of the orbiting flow that is receding along our sightline, producing partial absorption of the continuum at lower and/or redshifted projected velocities. On the other hand, even large rotational motion far from the accretion disk will only have a transverse velocity component within our line-of-sight column to the disk.
This argument against a rotational contribution to the line-of-sight velocity is supported by \citet{Borguet10}, who show that, for a spherically expanding outflow, a sharp absorption onset is only expected in cases with small rotational components. Any significant rotation results in a gradual absorption onset. 

This same grazing sightline in the H13 model also appears to reproduce the ramped blue edge of the C{\sc iv} absorption trough that we observe in H1413+117. The models of \citet{Giustini12} show that a ramped increase in transmission at the blue edge indicates a large variation in both velocity and density in the absorbing region along the line of sight, which appears to occur when this line of sight is close to tracing a streamline in the flow. This is consistent with a line of sight along the upper surface of the wind.

The interpretation of \citet{FilizAk14} contradicts this. They classify SDSS BAL quasars according to the appearance of Si{\sc iv} and Al{\sc iii} troughs alongside the ubiquitous C{\sc iv} trough, and interpret those containing all three (H1413+117 does, making it a LoBAL) as resulting from lines of sight on the underside of the outflow, intersecting the base of the wind.

The argument of the grazing sightline does not explain the different velocity shifts we see in the absorption onset for different line species, which should be the same if caused by our line of sight exiting the outflow. While ion species with the highest ionization potentials, C{\sc iv} (64.5~eV) and N{\sc v} (97.9~eV), have similar $\sim$1,500~km/s absorption onsets, the onset for Si{\sc iv} (45.1~eV) is $\sim$3,000~km/s and the deeper part of the Al{\sc iii} (29.4~eV) trough starts at $\sim$4,500-5,000~km/s (though may show weaker absorption from $\sim$2,000~km/s). 
This increase in absorption onset with diminishing ionization potential may suggest that for Si{\sc iv}  and Al{\sc iii} we are in fact seeing very sudden jumps in transition rate as ionizing radiation and/or gas density reach optimal levels as a function of radius and hence outflow velocity. For example, this may be the radius where it is first possible for ``Locally Optimized Clouds'' \citep{Baldwin95} to form due to the diminishing ionization parameter. 

A combination of the above may be needed to explain the onset of all absorption troughs; we propose that for C{\sc iv}  and the higher ionization N{\sc v} we are seeing our line of sight exit the outflow near its launch point, while for lower ionization lines we are seeing the onset of favorable absorption conditions further out in the flow. In the ionization/radiative transfer calculations of H13, high C{\sc iv} fractions are expected at the highest ionization parameters in their luminous benchmark quasar (M$_{BH} = 10^9$~M$_\odot$, Eddington ratio = 0.2), and so it is expected that C{\sc iv} absorption should be seen at its inner-most radii.

\subsection{Interpretation of Emission Line Spectrum}
\label{belrcomp}

Figures \ref{decomp} \& \ref{civdecomp} (lower) shows a plausible BEL spectrum, however it should be remembered that this component includes the BAL absorption of this broad line emission. It is not possible to extricate the true absorption profile ${\bf A}_L$ from the line profiles ${\bf F}_L$. In the case of the C{\sc iv} line (Fig. \ref{civdecomp}, lower) we see a region of near-zero emission in the rest frame 1515-1525\AA\ range ($\sim$4500--6400 km/s) where we might otherwise expect to see broad-line emission---either C{\sc iv} or Fe{\sc ii}, given the presence of one or both surrounding this region, and the significant iron emission in this wavelength range of the \citet{VW01} template. We take this as evidence of absorption of broad-line emission by the BAL wind. If this black region is due to absorption then we are seeing very high column densities {\em and} very high covering factors for absorption of the C{\sc iv} and/or Fe{\sc ii} emission at these velocities. This is very interesting due to the presumably much larger extent of the source of this broad-line emission compared to the continuum---the line-emitting gas is completely enveloped by absorbing gas across the entire BELR at these projected velocities.

Further evidence of absorption of BEL light by the BAL wind comes from the significant absorption of Ly$\alpha$ by the N{\sc v} trough in this source (see also Turnshek 1987).
While the symmetry of the C{\sc iv} emission in this component resembles that of an unabsorbed BEL, it should be remembered that C{\sc iv} very frequently shows a strong blue excess which may be absorbed in this case. The Si{\sc iv} line shows a sharp drop in flux in its blue wing that also suggests absorption of BEL light.

The iron emission template of \citet{VW01} shown in these figures provides a good match to several features in the emission line spectrum, and an imperfect match to others. The complexity and potential variation in iron emission makes a perfect match unlikely, however the scale and shape of these features match the template adequately. As well as natural variation, it is also possible that we are seeing differential microlensing acrosss the iron emission spectrum. The strong presence of iron features in component ${\mathcal L}$ implies that some of this emission arises on size scales similar to other BELs.

The fact that the iron template provides a reasonable match to the iron emission across the full wavelength range of the BELR component (Fig. \ref{decomp}) gives us confidence in our choice of continuum, and that we avoided most of the iron emission; if not then we would expect a significant and wavelength-dependent mismatch between the decomposed BEL spectrum and the template.

\subsection{Features in the C{\sc iv} and Si{\sc iv} Troughs}
\label{features}

Low-velocity-width emission peaks are seen within the region of the C{\sc iv} BAL absorption trough, marked in Figure \ref{civdecomp}: in the continuum component centered at blueshifts of $\sim$8,000~km/s ({\it a})) and $\sim$11,000~km/s ({\it b}) and in the BEL component at $\sim$4,000~km/s ({\it c}), $\sim$7,500~km/s ({\it d}), and $\sim$13,000~km/s ({\it e}).
An emission peak is also observed blueward of Si{\sc iv} $\lambda\lambda$1393.76 at $\sim$4,500~km/s (which we designate feature {\it f}, Fig.~\ref{pcygni}).

H10 observe a similar emission peak in their decomposition; it appears in all epochs in $\sim$6,500---9,500~km/s blueward of C{\sc iv}  in their ${\bf F}_{M}$ (BEL) component (from the decomposition of lensed image D with respect to A \& B). This appears to correspond to our feature {\it d}. Also notable is a feature in their ${\bf F}_{M\mu}$ (continuum) component from the decomposition of image C which corresponds to our feature {\it a}. A similar peak was also observed in the C{\sc iv} absorption profile derived by A90 at $\sim$5,500~km/s. In all of these previous decompositions, signal quality makes it is difficult to rule out the presence of any of the other features that we note in the current work.

The features observed in the C{\sc iv} and Si{\sc iv} troughs of the continuum component ({\it a}, {\it b} \& {\it f}) indicate regions of reduced absorption in the BAL outflow. This may be due to radial fluctuations in overall outflow density and/or relative ion species density due to variations in ionization parameter.
This interpretation reflects that of H10, who propose a strongly absorbing cloud embedded within the outflow but with lower velocity width that results in a double-peaked C{\sc iv} line.
In the context of a sightline grazing the upper layer of a fast, equitorial outflow, \citet{Giustini12} show that, in the hydrodynamic model of \citet{Proga04}, ``blobs'' of failed outflow can cause clumpiness in this layer as they fall from the over-ionized high-inclination regions. 

Interpretation of the features in the BEL component is more complicated. Features {\it d} and {\it e} may result from clumpy absorption of a strong blue excess in C{\sc iv} and/or Fe{\sc ii} emission (which may or may not be absorbed). As discussed in Section \ref{belrcomp}, the near-black region between features {\it c} and {\it d} does seem to indicate absorption of BEL light by the outflow. Whether this is also true of the dark region between features {\it d} and {\it e} is unclear, as feature {\it e} may be isolated Fe{\sc ii} emission. Feature {\it c} is probably iron emission.

It is clear that features {\it a} and {\it d} correspond to the same peak in the original spectrum, and so to the same region of diminished absorption. Feature {\it b} in the continuum component has no obvious counterpart in the BEL component. This may be due to intrinsically low BEL flux in this region or a difference in the BEL vs. continuum absorption profiles. 

\subsection{Line-of-Sight Differences}
\label{lineofsight}

The differences observed in the BAL profiles between lens images cannot be due to differences in the line of sight. The BAL absorption profile affecting the continuum is determined by the region of the BAL flow sampled by the column along the line of sight to the accretion disk, and so different lensed images will only experience different absorption if their line-of-sight columns diverge significantly within this outflow.
The image separation between images A and D is 1\farcs18, which corresponds to a divergence angle of the A and D lines of sight from the source of 0\farcs66 and 2\farcs0 for $z_L = 0.94$ and 1.4 respectively. 
Assuming $z_L = 1.4$ and a conservatively small accretion disk with half-light radius $10^{13}$ m (SS disk with $L/L_{EDD} = 0.1$ and $M_{BH}=10^8 M_\odot$), lines of sight have diverged by one half-light radius at $\sim$100 light years. This is probably much larger than the strongly-absorbing region of the BAL flow. Choosing a smaller $z_L$ gives an even smaller divergence.
On the $\lesssim$1000 light day size scale of a large BELR \citep{Kaspi00, Kaspi07}, line-of-sight columns to the accretion disk still have $\gtrsim 98$\% overlap.

Line-of-sight columns to the BELR, assuming 100 light day cross-section, diverge completely at $\sim$28,000 LY, and so may be subject to different line-of-sight absorption features both within the host galaxy and in the IGM. This is also true of the continuum, however the continuum will be subject to differences in absorption features from a greater depth within the host galaxy.

These line-of-sight differences do account for the many differences in narrow absorption features observed in our spectra. \citet{Monier99} show that these features arise both in the IGM and potentially the lensing galaxy, but probably not the host galaxy which has a higher redshift than observed features. We don't investigate the absorption features in detail in this paper, however we note that the $z = 1.43$ Fe{\sc ii} lines ($\lambda_{rest}$ 2344.2\AA, 2374.5\AA, 2382.8\AA) follow the same behavior in our spectra as observed by \citet{Monier99}; it is present in lensed images A and B and absent in C and D. Several other features exhibit similar behavior, and we will investigate these in a separate study.

\section{Conclusions}
\label{conc}

We have presented new spectroscopy of the gravitationally lensed broad absorption line quasar H1413+117 using Gemini North's GMOS IFU. Extracted spectra of the four lensed images reveal that differential microlensing strongly affects the relative magnification of the broad emission lines compared to the continuum in lensed image D, with the continuum in D appearing anomalously bright compared to all other lensed images. Lensed images A, B, and C also show small differences in their continuum-to-broad line flux ratios relative to each other.

Subtracting a fit to the continuum emission outside the broad emission and absorption regions allows us to determine the wavelength-dependent D/A flux ratios of the broad emission lines. These flux ratios are flat, indicating no differential microlensing across the velocity structure of the BELR.

Monte Carlo simulations were performed to analyze the dependence of the flatness of broad line flux ratios on the the size range spanned by the BELR velocity structure. It was found that BELRs with an upper size limit in the $10^{14}$---$10^{15}$~m range, and spanning 0.5 dex across their velocity structure are very unlikely to produce the observed flat flux ratios. Smaller BELRs are also unlikely due to the strong differential magnification observed between broad lines and continuum. Therefore, it is inferred that the BELR in H1413+117 is significantly larger than $10^{15}$~m, and produces no significant emission from regions smaller than $\sim 5\times10^{14}$~m, or 20 light days.

We take advantage of the differential magnification between lensed images A and D to algebraically decompose the quasar spectrum into two components: the absorbed broad line spectrum and the absorbed continuum spectrum, and we use the latter to derive intrinsic absorption profiles. Our decomposition leads us to these conclusions:

\noindent
$\bullet$ A sharp, blue-shifted onset of absorption is observed in all prominent BAL features in the continuum component. In the case of C{\sc iv}, N{\sc v}, and Ly$\alpha$ this onset is at $\sim$1,500~km/s, which may represent the inner edge of a radial outflow. In the context of typical disk-wind models this would place the onset of BAL absorption near its launch point, and so in the vicinity of the accretion disk. In the case of Si{\sc iv} and Al{\sc iii}, which have a larger absorption onsets, we may be seeing increases in absorption efficiency of lower ionization species with distance from the ionizing source. This sharp onset also indicates that the line-of-sight velocity is dominated by outflow, not rotational motion.

\noindent
$\bullet$ Overall the derived absorption profile for C{\sc iv} resembles that of an accelerating outflow that is either spherical or viewed relatively close to a streamline, albeit with a $\sim$1,500km/s initial velocity.

\noindent
$\bullet$ The C{\sc iv} and/or Fe{\sc ii} BELR appears to be strongly absorbed in the range $\sim$4,500---6,400~km/s. It is unclear whether emission features blueward of this trough are C{\sc iv} or Fe{\sc ii}, however the complete absorption of either of these features suggests a very high covering factor for the absorbing material at large radii. Combined with the apparent strong absorption near the outflow launch point, this suggests that gas responsible for BAL features maintains a high-covering-factor over at least two orders of magnitude in size.

\noindent
$\bullet$ We observe regions of reduced absorption within the BAL outflow, indicating clumpiness in gas density or relative ion species density.

\end{document}